\documentclass[amstex,12pt]{article}
\usepackage{amssymb}
\usepackage{cite}
\newlength{\dinwidth}
\newlength{\dinmargin}
\newtheorem{Definition}{Definition}[section]

\newtheorem{Proposition}[Definition]{Proposition}
\newtheorem{Lemma}[Definition]{Lemma}

\newenvironment{proof}{\medskip \noindent 
            {\bf Proof.}}{ \hfill $\square$ \medskip}

\renewcommand{\theequation}{\thesection.\arabic{equation}}
\def\ni{\noindent}
 
\def\be{\begin{equation}}
\def\ee{\end{equation}}

\def \NN {{\mathbb N}}    %natural
\def \RR {{\mathbb R}}    %real
\def \g {{\gamma}}

\def \ph {{\varphi}} %Achtung: Form haengt vom Textsystem ab

\def \tint{\textstyle \int}
\def\AO {\mbox{${\cal A}({\cal O})$}}
\def\AO'{\mbox{${\cal A}({\cal O}')$}}
\def\O {\mbox{${\cal O}$}}

\def\A{\mbox{${\cal A}$}}

\def \ijk {{\varepsilon_{ijk}}}

\def \O {{\cal O}}
\def \Op {{{\cal O}^{\,\prime}}}
\def \A {{\cal A}}
\def \AO {\A(\O)}
\def \AOl'{\A(\O_{loc}')}

\def \B {{\cal B}}

\def \D {{\cal D}}

\def \jm {{j_\mu}}

\def \jz {{j_0}}

\def \Fmn {{F_{\mu \nu}}}
\def \Fnm {{F_{\nu \mu}}}

\def \Am {{A_\mu}}
\def \Amu {{A^\mu}}

\def \An {{A_\nu}}
\def \Aj {{A_j}}

\def \Az {{A_0}}

\def \psb {{\bar\psi}}

\def \dm {{\partial_\mu}}

\def \dn {{\partial_\nu}}

\def \dz {{\partial_0}}
\def \dzu {{\partial^{\,0}}}

\def \dmu {{\partial^{\,\mu}}}
\def \dnu {{\partial^{\,\nu}}}

\def \A {{\mbox{\boldmath$A$}}}
\def \E {{\mbox{\boldmath$E$}}}
\def \B {{\mbox{\boldmath$B$}}}
\def \j {{\mbox{\boldmath$j$}}}

\def \x {{\mbox{\boldmath$x$}}}
\def \y {{\mbox{\boldmath$y$}}}
\def \z {{\mbox{\boldmath$z$}}}

\def \dmA {{\partial_\mu A^\mu}}

\def \gmy {{g_\mu (y)}}

\def \gmu {{\gamma^\mu}}
\def \gam {{\gamma_\mu}}

\def \Dm {{D_\mu}}
\def \Dmu {{D^\mu}}

\def \evp {(e/4\pi)\,}
\def \ovp {{(4 \pi)}^{-1}\,}

\def \roii {{\mbox{\scriptsize \it II}}}
\begin{document}
\title{Quantum Delocalization of the Electric Charge}
\author{Detlev Buchholz$^a$, \ Sergio Doplicher$^b$, \ 
Giovanni Morchio$^c$,\\ John E. Roberts$^d$,  
\ Franco Strocchi$^e$ \\[5mm]
\normalsize
$^a\,$Institut f\"ur Theoretische Physik der Universit\"at G\"ottingen,\\
\normalsize 
D-37073 G\"ottingen, Germany\\ 
\normalsize 
$^b\,$Dipartimento di Matematica, Universit\`a di Roma ``La Sapienza'',\\
\normalsize 
I-00185 Roma, Italy \\ 
\normalsize 
$^c\,$Dipartimento di Fisica dell'Universit\`a, I-56126 Pisa, Italy,\\ 
\normalsize 
$^d\,$Dipartimento di Matematica, Universit\`a di Roma ``Tor Vergata'',\\
\normalsize 
I-00133 Roma, Italy\\ 
\normalsize 
$^e\,$Scuola Normale Superiore, I-56126 Pisa, Italy}
\date{\today}
\maketitle
%%%%%%%%%%%%%%%%%%%%%%%%%%%%%%%%%%%%%%%%%%%%%%%%%%%%%%%%%%%%%%%%%%%%%%%%%%%%
%%%%%%%%%%%%%%%%%%%%%%%%%%%%%%%%%%%%%%%%%%%%%%%%%%%%%%%%%%%%%%%%%%%%%%%%%%%%
%%%%%%%%%%%%%%%%%%%%%%%%%%%% Abstract %%%%%%%%%%%%%%%%%%%%%%%%%%%%%%%%%%%%%%
%%%%%%%%%%%%%%%%%%%%%%%%%%%%%%%%%%%%%%%%%%%%%%%%%%%%%%%%%%%%%%%%%%%%%%%%%%%%
\begin{abstract} \noindent
The classical Maxwell--Dirac and Maxwell--Klein--Gordon
theories admit solutions of the field equations where the  
corresponding electric current vanishes in the causal
complement of some bounded region of Minkowski space. 
This poses the interesting question of whether states with 
a similarly well localized charge density also exist in quantum 
electrodynamics. For a large family of charged states, the dominant 
quantum corrections at spacelike infinity to the expectation 
values of local observables are computed. It turns out that 
certain moments of the charge density
decrease no faster than the Coulomb field in spacelike directions. 
In contrast to the classical theory, it is therefore  
impossible to define the electric charge support of 
these states in a meaningful way.
\end{abstract}

%%%%%%%%%%%%%%%%%%%%%%%%%%%%%%%%%%%%%%%%%%%%%%%%%%%%%%%%%%%%%%%%%%%%%%%%%%%%
%%%%%%%%%%%%%%%%%%%%%%%%%%%%%%%%%%%%%%%%%%%%%%%%%%%%%%%%%%%%%%%%%%%%%%%%%%%%
%%%%%%%%%%%%%%%%%%%%%%%%%%%%%%%%% SECTION 1 %%%%%%%%%%%%%%%%%%%%%%%%%%%%%%%%
%%%%%%%%%%%%%%%%%%%%%%%%%%%%%%%%%%%%%%%%%%%%%%%%%%%%%%%%%%%%%%%%%%%%%%%%%%%%
\section{Introduction}
Electrically charged systems are known to have  
poor localization properties with regard to measurements of the electric 
field, which extends to spacelike infinity    
according to Coulomb's law. Whereas  
this delocalization is an inevitable consequence of 
Maxwell's equations, 
the electron comes close to the idea of a point 
particle and one might infer 
that the charge density of such systems ought to have much
better localization properties. Thus the 
interesting question arises of  
whether it is possible to assign a charge support in a consistent manner.

The idea of a clearcut distinction between charge and field 
support seems unproblematic in the context of classical physics.
The possibility that 
it might be meaningful in quantum field theory too was first 
considered by Fr\"ohlich \cite{Fr}
in a general discussion of the superselection
structure of electrically charged states. More recently, 
a simple non--interacting 
model allowing a precise definition of the electric 
charge support of states was presented in \cite{BuDoMoRoSt}, 
where it was also outlined how this notion could be used 
to analyze the statistics and symmetry properties
of such models by generalizing methods developed 
in \cite{DoHaRo, DoRo} for analyzing strictly localized states. Thus  
a systematic investigation of the localization properties of the electric 
charge in more realistic (interacting) theories seems warranted.

It is the aim of the present article to carry out such an analysis
in classical and quantum electrodynamics. In the classical  
Maxwell--Klein--Gordon and Maxwell--Dirac theories it turns out 
that a charge support can be sharply defined.
More precisely, there exist finite energy solutions of the 
coupled field equations such that 
the corresponding electric current vanishes in the causal complement  
of some double cone in Minkowski space whereas the 
electric field is of Coulomb type there. These results fit 
perfectly with the heuristic picture of a point--like support 
of the electric charge.

The simple picture breaks down, however, if one takes quantum
effects into account. Using perturbative methods, we shall determine,
for a large family of charged physical states
in quantum electrodynamics, the dominant quantum 
contributions to the matrix elements of local observables
at spacelike infinity.
These contributions have a simple form 
and thus can be computed in the cases of interest: 
whereas the expectation values of the 
charge density and their mean square fluctuations 
exhibit spatial decay properties which seem to 
corroborate the picture of a reasonably well localized charge 
distribution, the  higher moments decay no faster
than the Coulomb field. As a matter of fact, these moments can be used
to determine the shape of the asymptotic electromagnetic field
of the states. So the idea of discriminating the charge support 
from the field support fails in these examples.

The origin of this phenomenon will be traced back to vacuum 
polarization effects, namely the fact that observables which are
related to the matter fields can generate states from the vacuum  
containing, with non--zero probability, only low energy photons. If 
the interaction is turned off, this effect disappears and the 
resulting states have a mass gap. This general mechanism is
also at the root of a result by Swieca \cite{Sw} who proved that 
the spatial integral of the charge density in electrically
charged states exhibits an oscillatory behaviour
in time, thereby leading to a Coulomb--like delocalization 
of the spatial components of the electric current. 
The present results show that the delocalizing effects of  
vacuum polarization also affect the charge density. Although 
we restrict attention here to a special family of charged states 
(corresponding to gauges of ``Coulomb type''), our results 
provide evidence to the effect that this delocalization is generic. 

Our paper is organized as follows. In Section 2 we
use global existence theorems for the   
classical Maxwell--Klein--Gordon and Maxwell--Dirac theory  
to  exhibit solutions of the field equations with 
sharp support properties of the electric current.
The quantum induced delocalization of electrically 
charged states is discussed in Section 3.
By slightly modifying a perturbative method for constructing 
charged states, 
established by Steinmann \cite{St}, we 
exhibit a family of such states where  
the asymptotically leading quantum corrections to 
the matrix elements of local observables 
can conveniently be analyzed. The paper concludes with a brief
discussion of the implications of our 
results for the general analysis of the superselection structure of 
theories involving charges of electric type. 
%%%%%%%%%%%%%%%%%%%%%%%%%%%%%%%%%%%%%%%%%%%%%%%%%%%%%%%%%%%%%%%%%%%%%%%%%%%%
%%%%%%%%%%%%%%%%%%%%%%%%%%%%%%%%% SECTION 2 %%%%%%%%%%%%%%%%%%%%%%%%%%%%%%%%
%%%%%%%%%%%%%%%%%%%%%%%%%%%%%%%%%%%%%%%%%%%%%%%%%%%%%%%%%%%%%%%%%%%%%%%%%%%%
\section{Localizing the classical electric charge}
\setcounter{equation}{0}

We begin by discussing the localization properties of the electric
charge at the classical level in the Maxwell--Klein--Gordon 
and Maxwell--Dirac theory. 
In the Maxwell--Klein--Gordon theory, the field equations are 
\be 
\dnu \Fmn = \jm, \quad \Dm \Dmu \ph = 0.
\ee
Here $\Fnm = \partial_\nu \Am - \partial_\mu \An$ 
is the electromagnetic field, $\Am$ 
is the vector potential in the Coulomb gauge,
$\ph$ is the charged scalar field and 
$\Dm \equiv \dm - ie\Am$, where $e$ is the unit of charge. The 
electric current $\jm$ is given by 
\be
\jm = e \, \Am \, \ph^* \ph - e \, {\textstyle {i\over 2}} \, 
(\ph^* \dm \ph - \dm \ph^* \, \! \ph). 
\ee
By using the existence theorem of Klainermann and Machedon \cite{KlMa},
we shall exhibit solutions of these equations 
where, in a given Lorentz system, both the 
current $\jm$ and the magnetic field 
$B_i \equiv {1 \over 2} \, \ijk \, F_{j k}$ 
vanish in the spacelike complement of some compact
region, whereas the electric field $E_i \equiv F_{0\, i}$ is 
non--zero there, $i = 1,2,3$.

\begin{Proposition}
In the classical Maxwell--Klein--Gordon theory, for any double
cone $\O \subset \RR^4$, there are finite energy solutions $\ph$, $\Am$
with non-zero charge,  
\be
\tint d\x  \, \jz (\x , x_0) \neq 0, 
\ee
such that for any $x = (\x,x_0) \in \Op$
\begin{eqnarray} 
& \jm (x) = 0, \ \B (x) = 0  &  \label{2.5}  \\
& \E (x) = \ovp
{\tint} d\y \, (\x - \y) \, 
|\x - \y|^{-3} \, \jz (\y , 0). \label{2.6} & 
\end{eqnarray}
\end{Proposition}
\begin{proof} By the main theorem of Klainermann and Machedon \cite{KlMa},
the Maxwell-Klein--Gordon equations in the Coulomb gauge have  unique 
global solutions with finite energy for all smooth initial data
$ \Aj,\, \dz\Aj,\,\,  \ph,\,  \dz \ph\,$ with
support in the base of $\O$. Moreover, these solutions are smooth 
in all variables. We shall show that their support properties are 
as stated above.

To see this, we note that the field equation for  $\ph$  can be
read as a hyperbolic equation in the ``external'' field $\Am$, 
$$
\square \ph  = -2 ie
\Am \dmu \ph + e^2\Am A^\mu \ph + ie \, \dzu\Az \, \ph
 \equiv J (\ph, \dm \ph).
$$
Setting $\chi \equiv (\ph, \dz \ph)$ gives a first order system 
satisfying the following integral equation with respect to the time variable  
\begin{eqnarray*}
& \chi(x_0) = G(x_0) \chi(0) + 
{\displaystyle \tint_0^{x_0}} dy_0 \, G(x_0-y_0) f(\chi(y_0)). & \\ 
& f(\chi(y_0)) \equiv (\, 0 \, , \, J (\ph, \dm \ph )(y_0)), &
\end{eqnarray*} 
Here $G(x_0)$ is the Green's function 
(propagator) of the equation for $e=0$ and the 
spatial dependence has been suppressed. 
This integral equation is known to have a unique solution $\chi$, provided 
$f$ satisfies a suitable local Lipschitz condition,
see e.g.~\cite{PaStVe}. 
In our case, this Lipschitz condition holds because $\Am$ is 
smooth, and therefore locally bounded.

Because of the hyperbolic character of the equation, the 
solution in a given double cone depends only on the initial 
data on the base of that double cone. 
Thus, for initial data $\ph , \dz \ph$ of compact support
contained in $\O$, $\ph$ and therefore $\jm$ vanish in the
causal complement $\Op$. Moreover, Maxwell's equations give 
$$
\square \B =  \mbox{curl}\, \j, 
$$
and therefore $\B$ vanishes in $\Op$ 
if the initial data for $\A$ have support
in $\O$. Finally, by Maxwell's equations,
$${\dz \E = \mbox{curl} \, \B - \j = 0 \ \ \mbox{in} \ \Op,}$$
so $\E$ is time independent in $\Op$ and therefore given by
equation (\ref{2.6}).
\end{proof}

Let us now turn to the Maxwell--Dirac theory with the field equations 
\begin{eqnarray}
& \dnu \Fmn  =  \jm  =  e \, \psb \gam \psi, & \nonumber \\
& ( - i \gmu \dm + m ) \, \psi  =  e \, \gmu \Am  \, \psi, & 
\end{eqnarray} 
where $\psi$ is the Dirac field and $\gmu$ are the gamma matrices. 
Here the results are slightly weaker than in the preceding case
since the initial value problem is under control only
for small initial data.

\begin{Proposition} In the classical
Maxwell--Dirac theory, given any double cone $\O$, there are finite
energy solutions $\psi$, $\Am$ with (small) non--zero charge $q$, 
\be
\tint d\x \, \jz (\x , x_0) = q, 
\ee
such that the corresponding current and the electromagnetic 
field have in $\Op$ the properties (\ref{2.5}), (\ref{2.6}).
\end{Proposition}
\begin{proof} Theorem 2.5 of \cite{FlSiTa1} (see also \cite{FlSiTa2})
establishes the existence and uniqueness of finite energy  
solutions for sufficiently small smooth initial data. 
In particular, if the initial data for $\psi$ and $\A$ are 
smooth, have compact support and are bounded by a sufficiently 
small constant and the initial data for $A_0$ are computed from the 
gauge condition 
$$
\dmA = 0
$$
and the Gauss constraint 
$$
\Delta \,A_0= - \partial^i \,\dz\,A_i + e \, |\psi|^2, 
$$
a unique global solution exists 
and satisfies the preceding gauge conditions at all times. 
Moreover, this solution is smooth in the spatial variables
and locally bounded in time.
Thus, with the global Cauchy problem for the Maxwell--Dirac equations 
for sufficiently small smooth data (corresponding to small electric 
charge) under control, we can proceed as in the scalar case. 
The solution $\psi$ solves the
integral equation 
$$
\psi(x_0)=S(x_0)\,\psi(0) + i e
\tint_0^{x_0} dy_0 \, S(x_0-y_0) 
\g_0 \gmu\,\Am(y_0)\,\psi(y_0), 
$$
where 
$S(x_0)$ is the Green's function (propagator) of the free Dirac 
equation which has 
the same hyperbolic properties as $G(x_0)$ in the scalar case and
$\Am$ is regarded as an external field. Thus
the non--linear term again satisfies a local Lipshitz condition 
since $\Am(\x,x_0)$ is bounded in $\x$ uniformly for
$x_0$ in finite intervals. By the same argument as in the scalar
case, therefore, $\psi$ vanishes in $\Op$. This implies
that $\jm$ vanishes in $\Op$, and the results for $\B$ and $\E$
follow as before.
\end{proof}

Thus we find that, in classical
field theory, the localization
properties of the electric charge are not affected  
by the interaction between the  
electromagnetic field and the matter fields.
%%%%%%%%%%%%%%%%%%%%%%%%%%%%%%%%%%%%%%%%%%%%%%%%%%%%%%%%%%%%%%%%%%%%%%%%%%%%
%%%%%%%%%%%%%%%%%%%%%%%%%%%%%%%%%%%%%%%%%%%%%%%%%%%%%%%%%%%%%%%%%%%%%%%%%%%%
%%%%%%%%%%%%%%%%%%%%%%%%%%%%%%%%% SECTION 3 %%%%%%%%%%%%%%%%%%%%%%%%%%%%%%%%
%%%%%%%%%%%%%%%%%%%%%%%%%%%%%%%%%%%%%%%%%%%%%%%%%%%%%%%%%%%%%%%%%%%%%%%%%%%%
\section{Quantum delocalization} 
\setcounter{equation}{0}

We want to study now how quantum effects modify the
localization properties of the electric charge. As a rigorous
construction of quantum electrodynamics has not yet been 
accomplished, we have to rely on perturbative methods 
and results in this analysis.

Here, it is convenient to use the (indefinite metric) Gupta--Bleuler
formalism of quantum electrodynamics  based on the (unphysical)
local Dirac field $\psi$ and the local vector potential $\Am$.  
The existence of the corresponding renormalized
Green's functions has been established to all orders in perturbation theory 
by various methods to a
by now satisfactory degree of rigour \cite{BlSe,FeHuRoWr,KeKo}.  

The problem of constructing physical charged fields and states in the 
Gupta--Bleuler formalism, however, requires further 
analysis. As first pointed out by Dirac, such fields can 
be obtained by formally multiplying 
$\psi$ with non--local operators which restore the local 
gauge invariance,
\begin{equation} \label{3.1} 
\psi(x) \, \exp{\big( ie A(g_x) \big)},\end{equation} where 
\begin{equation} A(g_x)\equiv 
\tint dy\,\Amu(y)\,g_{x \, \mu}(y) \quad \mbox{and} \quad {\dmu
g_{x \, \mu}(y)=-\delta(y-x).} \label{3.2} \end{equation} 
The rigorous treatment of these expressions requires 
control both of infrared and of ultraviolet problems.
The infrared problems appear because of the slow decay of 
the ``gauge fixing functions'' $g_{x \, \mu}$
and the ultraviolet problems are due to the singular nature of
the products of field operators involved in the definition of 
the exponential $\exp{\big( ie A(g_x) \big)}$ (in the sense of 
formal power series) and of $\psi(x)$. 
These problems have been extensively discussed  
by Steinmann \cite{St}, cf.\ also \cite{Sy}, who established 
the existence of physical charged fields
for a large class of gauge fixing functions within the framework of 
perturbation theory.
We will rely here on these results and manipulate
formal expressions such as (\ref{3.1}) freely, without
going into the subtle details of their precise definition.

In our analysis we also make use of the following general properties 
of the Gupta--Bleuler formulation of quantum electrodynamics
which have been established 
in perturbation theory. In order to keep the notation simple, 
we deal in the following 
with the unregularized fields $\psi(x)$, $\Am(y)$ etc. The
subsequent statements are thus to be understood in the sense of 
distributions.\\[2mm]
1) The Wightman functions (vacuum expectation values) of the renormalized 
fields $\psi, \, \Am$ exist as tempered distributions satisfying
locality, Poincar\'{e} covariance and the spectrum 
condition \cite{BlSe,Sch}. \\[2mm]
2) The field $\dmA$ is the generator of c--number gauge
transformations in the sense that \cite{Sy,StWi} 
\begin{eqnarray} & [\dmA(z),\,\psi(x)] = e
    D(z-x)\,\psi(x), & \label{3.3} \\
& [\dmA(z),\,\An(y)] =-i \, \dn \, D(z-y), & \label{3.4}
\end{eqnarray}
where $D$ is the Pauli--Jordan distribution. The latter equation implies 
\be \label{3.5}
[ \dmA (z), \exp{(ie A(g))} ] = e \! \int \! dy \, D(z - y) \, 
\partial^\mu g_\mu (y) \cdot \exp{(ie A(g))}
\ee
for arbitrary $g_\mu$.\\[2mm]
3) Polynomials in the fields $\psi(x), \Am(y)$
commuting with $\dmA(z)$ are elements of the algebra of observables and    
have vacuum expectation values satisfying 
Wightman positivity in the sense of formal power series in $e$, cf.\  
\cite{DuFr}. As a consequence, these expectation values have the 
same cluster properties as in a positive metric Wightman field theory.

\vspace*{2mm}
We begin our analysis by explaining 
how one can proceed from the charged fields (\ref{3.1}) 
to unitary charge carrying operators. This step relies 
on the following two observations. 

First, as shown
in the appendix, for given 
$R$, $T$ and $x$ varying in the bounded 
region ${\cal R} \equiv \{ y : |\y| < R, |y_0| < T \}$ 
there exist gauge fixing functions $g_{x \, \mu}$ 
decomposable as  $g_{x \, \mu} = g_{x \, \mu}^{I} + 
g_{\mu}^{\roii}$, where $g_{x \, \mu}^{I}$, $g_{\mu}^{\roii}$
have the following specific properties. The functions 
$y \rightarrow g_{x \, \mu}^{I} (y)$ have compact support in the region 
$|\y| < 4 R$, $|y_0| < T$ and are, for $|\y| < R$, given by
\be  g_{x \, 0}^{I}(y) \equiv 0, \quad \ g_{x \, i}^{I} (y) 
\equiv - \ovp \delta (y_0 - x_0)
(y_i - x_i) |\y - \x|^{-3}, \ \ i=1,2,3. 
\label{3.6} 
\ee 
Thus they coincide in the latter region with the gauge fixing functions 
considered by Steinmann \cite{St}. The functions $y \rightarrow
g_{\mu}^{\roii} (y)$, on the other hand, are given by
\be \label{3.7}
 g_0^{\roii}(y) \equiv 0, \quad \
g_i^{\roii}(y) \equiv  - \ovp  y_i \, |\y |^{-3} \, l(\y) \, h(y_0), \ \   
i=1,2,3,
\ee
where $h$ is a  test function with support in $(-T,T)$, 
$\int \! d y_0 \, h(y_0) = 1$, and $l$ is a smooth function
which is equal to $0$ for $|\y| < 3R$ and equal to $1$ for
$|\y| > 4R$. Thus the functions $g_{\mu}^{\roii}$ do not depend on the 
choice of $x$ within the above limitations. 

Second, it follows from the properties of $g_{x \, \mu}$ and 
the commutation relations (\ref{3.3}), (\ref{3.5})
that the slightly modified charged fields
\be \label{3.8}
\psi (x) \, \exp{\big( ie A(g_x^I) \big)} \, 
\exp{\big( ie A(g_{}^\roii) \big)}, \ \ x \in {\cal R},
\ee
commute with $\dmA (z)$ 
and consequently create charged physical states from the vacuum.
The advantage of these fields is that 
the non--local effects, needed to describe the asymptotic
Coulomb field, are clearly separated 
from the local effects encoded in $\psi$.

We proceed from the fields 
(\ref{3.8}) to (unbounded) operators by integrating them 
with test functions $f$ with support in $\cal R$.
As $g_\mu^\roii$ does not depend on $x$, this integration affects 
only the first two factors in 
the product (\ref{3.8}). Moreover, since the functions 
$g_{x \, \mu}^I$ are well--behaved  extensions of the 
gauge fixing functions (\ref{3.5}) considered by Steinmann \cite{St}, 
the ultraviolet problems involved in defining the product
could be controlled by similar methods. We therefore 
anticipate that the expression 
\be
\chi(f) \equiv \int \! dx \, f(x) \, \psi (x) \, \exp{\big( ie A(g_x^I) \big)} 
\ee
is meaningful and 
defines a closable operator which is localized in some bounded
space--time region fixed by the support properties of 
$f$ and $g_{x \, \mu}^I$. 

Making use of the commutation relations (\ref{3.3}), (\ref{3.5})
and the specific properties of
$g_{x \, \mu} = g_{x \, \mu}^{I} + g_{\mu}^{\roii}$, one 
finds that the local operators $\chi(f)^* \, \chi(f)$ and 
$\chi(f) \, \chi(f)^*$ commute with $\dmA (z)$ and therefore are
observables. Skipping some technical details, it 
follows that the partial isometry  
$V_f$ appearing in the polar decomposition $\chi(f) = V_f \, |\chi(f)|$
is a local operator transforming under gauge
transformations in the same way as $\chi(f)$. Moreover, multiplying $\chi(f)$
from the left and right with suitable elements of the 
algebra of local observables, one
can always arrange that $V_f$ be unitary \cite{Bo}.  

The preceding arguments thus provide evidence to the effect that 
in quantum electrodynamics unitary operators of the form 
\be \label{3.U}
U_f \equiv V_f \, \exp{(ie A(g^\roii))}
\ee
exist 
which are invariant under local gauge transformations and carry 
electric charge. These operators have the interesting 
feature that the $f$--dependent contributions of the Dirac  
field $\psi$ are completely absorbed in the local operators $V_f$,
whereas the properties of the corresponding asymptotic electromagnetic
field are encoded in the $f$--independent operator 
$\exp{(ie A(g^\roii))}$. As we shall see, this clearcut separation of the local
and asymptotic features of the charge carrying operators 
greatly simplifies the analysis of the corresponding charged states. 

Keeping $f$ fixed in the following, 
we define with the help of the unitaries $U_f$ the maps
\be
C \rightarrow \rho_f (C) \equiv U_f^{-1} \, C \, U_f^{ },
\ee
where $C$ are arbitrary local observables. If $C$ is localized
in the region $\O_r + \x$, where $\O_r$ denotes the 
double cone of radius $r$ centred at $0$ and $\x$ is a 
sufficiently large spatial translation, then $C$ and $V_f$ commute
by locality. Thus in this case
$$
\rho_f (C) = \exp{(- ieA(g^\roii))} \, C \, \exp{(ieA(g^\roii))} 
$$
\be \label{3.12}
= \sum_{k=0}^\infty \frac{(- i e)^k}{k!} \,
(\mbox{Ad}\, A (g^\roii))^k  (C) 
\ee
holds in the sense of formal power series. In the following lemma we 
analyze the action of the approximants $\rho^{(n)}$, given by the first
$n$ terms in this series, on observables $C$ localized in 
spacelike asymptotic regions. As the fields are unbounded, 
our results are to be understood 
in the sense of sesquilinear forms on 
$\D \times \D$, where $\D$ is the linear span of 
vectors obtained by applying locally regularized gauge--invariant polynomials 
$C^\prime$ in the fields to the vacuum vector
$\Omega$.

\begin{Lemma}
Let $C$ be any local observable, let 
$n \in {\NN}$, and let $t > 0$. Then, 
for $|x_0| < t$ and large $|\x|$,    
\begin{equation} \label{3.13}
\rho^{(n)}(C(x))
= C(x) + i e \,a_C (x) \cdot {1} + \, R_{C}(x) \end{equation}
on $\D \times \D$. Here 
\begin{equation} \label{3.14}
a_C(x) \equiv 
\ovp \! \int \! dz \,
z_j |\z|^{-3} h(z_0) \, ( \Omega, [ A^j (z) , C(x)], \, 
\Omega ) \end{equation} 
with $h$ as in (\ref{3.7}), 
and the matrix elements of the remainder 
$R_{C}(x)$ decrease at least like $|\x|^{-4}$, uniformly
in $|x_0| < t$. 
\end{Lemma} 
\begin{proof} The leading term 
$C(x)$ in the asymptotic expansion of $\rho^{(n)}(C(x))$
corresponds to the $k=0$ contribution to the series
(\ref{3.12}). The next term is given by 
the vacuum expectation value of the 
$k=1$ contribution which has the form 
$$ 
-ie \! \int \! dz \, g_\mu^\roii (z) \, ( \Omega, \, 
[\Amu (z),
C (x) ] \, \Omega ).  
$$
Plugging into the integral the expression given in (\ref{3.7}) 
and taking locality into account, one obtains, for
sufficiently large translations $\x$,  
the function $ie \, a_C$ appearing 
in the statement. 

The proof that the remainder 
$$ R_{C}(x) = \rho^{(n)}(C(x))
- C (x) - i e \,a_C(x) \cdot {1} $$
has the stated decay properties requires more work.
We begin by noting that if $C$ is localized in the
double cone $\O_r$, the multiple commutators 
$(\mbox{Ad}\, A (g^\roii))^k \, (C (x)) $ contributing to 
$\rho^{(n)}(C(x))$
are localized in $\O_{r + k T+ |x_0|} + \x$, as a consequence of 
the support properties of $g_\mu^\roii$ and locality.
So, in view of the 
spacelike commutativity of local gauge--invariant polynomials 
in the fields, it suffices to establish the 
asymptotic decay properties 
of matrix elements of  $R_{C}(x)$ 
between vectors of the form 
${C^\prime} \Omega$ and the vacuum $\Omega$. 
Next, we introduce the notation 
$$
A_h^j (\z) \equiv \int \! dz_0 \, h(z_0) \, A^j (\z,z_0)
$$
and recall that, as a consequence of temperedness and the spectrum
condition, it suffices to regularize the fields in 
the time variable in order to obtain operators 
depending smoothly on the spatial variables on
their natural domain of definition \cite{Bo2}.

Now for large $\x$ as above, the contribution arising from the $k=1$ term 
in  $R_{C}(x) $ has the form 
\begin{eqnarray*} 
\lefteqn{ ( C^\prime_0 \Omega , 
[ A (g^\roii) , C(x) ] \, \Omega )} & \nonumber \\ 
&  & =  \ovp \! \int \! d\z \,  z_j \, |\z|^{-3} \,  
( C^\prime_0  \Omega,  [A_h^j ( \z), C (x) ] \, \Omega ) \\ 
&  & = 
\ovp \! \int \! d\z \,  (z_j + x_j) |\z + \x|^{-3} \, 
( C^\prime_0 \Omega, \, [A_h^j (z + \x),
C (x) ] \, \Omega ), 
\end{eqnarray*}
apart from a factor $ie$.
Here ${C^\prime}$ has been replaced by 
$C^\prime_0 \equiv C^\prime - (\Omega, \, 
{C^\prime}  \Omega) \cdot 1$
since the vacuum expectation value of the commutator has been
subtracted in $R_{C}(x) $.
The second equality is obtained by substituting  
$\z \rightarrow \z + \x$, which is legitimate in the 
present setting since the matrix element 
under the integral is 
continuous in all variables.

Because of locality, 
the latter integral extends over a bounded region 
${\cal K} \subset \RR^3$ which can be held fixed for $|x_0| < t$ and 
$\x \in {\RR}^3$. Moreover, for $|x_0| < t$ 
and $\z \in {\RR}^3$ the operators
$[A_h^j (\z), C (x_0) ]$ are localized in the 
fixed double cone $\O_{r + T + t}$ and are
gauge invariant, like ${C}_0^\prime $. So we can apply 
the Araki--Hepp--Ruelle cluster theorem \cite{ArHeRu}
to their vacuum expectation values, 
cf.\  the properties of the Gupta--Bleuler formalism stated above.
Thus 
$$|( C^\prime_0  \Omega, \, [A_h^j (\z + \x),
C (x) ] \, \Omega )| < c_j \, |\x|^{-2}, $$
uniformly in $|x_0| < t$ and $\z \in {\RR}^3$. 
Combining this estimate with the preceding information, we obtain
the bound 
$$
{|( C^\prime_0 \Omega , [ A (g^\roii) , C (x) ] \, \Omega )|}  
\leq c^{\prime}_j \int_{\cal K} \! d\z \,
|z_j + x_j| \, |\z+\x|^{-3} \,
|\x|^{-2} \leq c^{\prime \prime}  
|\x|^{-4}. 
$$

The higher order terms ($k \geq 2$) can be treated similarly.
In fact, for $|x_0| < t$, $\x \in {\RR}^3$ we have 
\begin{eqnarray*} 
\lefteqn{| ( C^\prime \Omega , [A (g^\roii) 
, \ldots [A (g^\roii) , C (x) ] \ldots ] \Omega )|} \\ 
& \leq & {(4 \pi)}^{-k} \int_{{\cal K}} d\z_1 
\, |z_{j_1} + x_{j_1}| \, |\z_1 + \x|^{-3} 
\ldots \int_{{\cal K}} d\z_k 
|z_{j_k} + x_{j_k}| \, |\z_k + \x|^{-3} \times \\
& \times & \big| 
( {C^\prime} \Omega , [A_{j_1}^h (\z_1+\x), 
\ldots [A_{j_k}^h (\z_{k}+\x) , C(x) ] \ldots ] \Omega ) \big|, 
\end{eqnarray*}
where ${\cal K} \subset {\RR}^3$ is some fixed compact set.  
The multiple commutator function
is bounded in $\z_1, \ldots
\z_k \in {\RR}^3$, uniformly in $|x_0| < t$ and $\x \in {\RR}^3$.  
Thus the integral is bounded by $c \cdot |\x|^{-2k}$, 
completing the proof of the statement. 
\end{proof}

It is important here that the form of the leading terms of the
asymptotic expansion given in the preceding lemma does not depend on
the order $n >1$ of the approximants $\rho^{(n)}$ of the map $\rho_f$.
This fact allows us to establish the following statement on the spacelike 
asymptotic properties of the charged states $U_f \, \Omega$.

\begin{Proposition} Let $C$ be any local observable and
let $t > 0$. Then for large $|\x|$
in any order of perturbation theory, 
\begin{eqnarray}
\lefteqn{(U_f \, \Omega, C(x)\, U_f \, \Omega)} & & \nonumber \\ 
& = & (\Omega, C \, \Omega) +  i \evp \,  x_j \, |\x|^{-3} 
\, \int \! dz \, h(z_0)  
( \Omega, \, [ A_j (z) , C(x_0) ] \, \Omega )
\end{eqnarray}
apart from terms which decrease at least like $|\x|^{-3}$, uniformly
in  $|x_0|<t$.
\end{Proposition} 
\begin{proof} As  
$(U_f \, \Omega, C(x)\, U_f \, \Omega) =
(\Omega, \rho_f(C(x)) \, \Omega)$ 
and $\rho_f$ can be replaced in any given order
of perturbation theory by $\rho^{(n)}$ for sufficiently large $n$,
the statement follows from the preceding lemma by extracting the
asymptotically leading contribution in $|\x|$ from the function $a_C$
appearing there.
\end{proof}

Thus in the charged states, besides
the leading vacuum contribution of the observable $C$, 
a term appears behaving asymptotically like the Coulomb field,
whenever the corresponding integral does not vanish. 
In view of the commutator appearing in this expression, we call this 
sub--leading contribution the asymptotic quantum correction, for short.

It is instructive to study the form of this contribution for specific 
observables. Making use of locality, Lorentz 
covariance and the spectrum condition, it follows
that the commutator function of the vector potential has the form 
\be \label{3.15}
(\Omega, [ A_\mu (u) ,
A_\nu (v) ] \, \Omega ) = -i g_{\mu \nu} K(u-v) - i 
\partial_\mu \partial_\nu L(u-v), 
\ee
where $K$, $L$ are causal, Lorentz invariant distributions
whose Fourier transforms have supports in the forward and backward
light cones. 
Hence the expectation value of the electric field 
in the charged states has the asymptotic form
\be \label{3.16}
(U_f \, \Omega, \E(x)\, U_f \, \Omega) =
- \evp \x \, |\x|^{-3} 
\, \int \! dz \, h(z_0 + x_0) \, \partial_0 K (z),
\ee
in agreement with the expected Coulomb behaviour.  But, in contrast
to the situation in classical field theory, the Coulomb field is
modulated by an additional time dependent factor: 
only if $K$ is equal to the 
massless Pauli--Jordan commutator function, i.e.\ in zeroth order
perturbation theory, does this factor equal~$1$. Higher order 
(loop) corrections induce an additional oscillatory behaviour \cite{Sw} 
which may be attributed to vacuum polarization (i.e.\ quantum) effects 
which interfere with the asymptotic Coulomb--like contributions of the 
states $U_f \, \Omega$.
We emphasize that these oscillations are not in 
conflict with the fact that the charge of the underlying
state is equal to $e$. Determining the total 
charge of a state from the expectation values of the charge density  
requires in general not only an integration
over all of space but also a suitable 
mean over time \cite{Re}. Such a procedure also works  
for electrically charged states \cite{MoSt}. 
It is in fact easily checked that, as a consequence of 
Eqs.\ (\ref{3.4}) and (\ref{3.15}), the mean of the time dependent 
factor in relation (\ref{3.16}) is equal to $1$, in agreement
with the charge content of the underlying state.

If $C$ is the sum of spatial derivatives 
of local operators, the integral appearing in the 
asymptotic quantum correction vanishes. As $j_0 = \mbox{div} \E$
and $\B = \mbox{curl} \A$, we conclude that    
the matrix elements of the charge density and of the magnetic field
decrease at least like $|\x|^{-3}$. As a matter of fact,
a more refined analysis shows that they behave 
like $|\x|^{-6}$. Moreover, since the vacuum expectation 
values of triple products of the electromagnetic field vanish, 
as a consequence of the charge conjugation symmetry, the mean square 
fluctuations of the charge density in the charged states 
coincide asymptotically with those in the vacuum, up to 
contributions which decrease like $|\x|^{-4}$. So, 
in this sense, the charged states have localization properties with 
respect to these observables coming close to those in the
classical theory. 

A first clear deviation from the classical situation 
appears in the case of the spatial components of the current. 
Whereas classically these components have the same support 
properties as the charge density, one gets in the quantum case 
\be \label{3.17}
(U_f \, \Omega, \j(x)\, U_f \, \Omega) =
\evp \x \, |\x|^{-3} 
 \int \! dz \, h(z_0 + x_0) \, \partial_0^{\,2} K (z)
\ee
as the leading contribution. So these expectation values 
decrease asymptotically no faster than 
the Coulomb field. This result is related to 
the temporal oscillations of the electric field, mentioned above, 
and shows that quantum effects lead to a substantial  
asymptotic delocalization. 

It is {\it a priori} not clear whether quantum corrections 
and a corresponding asymptotic Coulomb like behaviour can also  
appear for higher moments of the better behaved observables, 
such as the charge density and the magnetic field. As a matter of 
fact, in spite of the restriction on the energy--momentum transfer
of the vector potential by the spacetime integration in the
asymptotic expansion given in Proposition 3.2, these
expressions are not controlled by general low energy theorems
\cite{JaRo}. The question can be decided, however, by perturbative 
calculations where one finds that, at one loop level, the asymptotic 
quantum corrections do not vanish for 
the  triple product $C = j_0(x_1) \, j_0(x_2) \, j_0(x_3)$,
even after averaging over time. We are indebted to O.\ Steinmann and 
O.\ Tarasov for communicating to us the results of 
these perturbative calculations. 

Recalling that for full information on a quantum observable 
all of its moments are needed, these results show that the charged 
states have no better localization properties 
with respect to measurements of the charge density
than the Coulomb field (although the amplitudes of the 
delocalizing terms are \mbox{suppressed} by
powers of the fine structure constant). Thus a meaningful separation 
between the charge and field support of these states 
is impossible, in contrast 
to the simple model considered in \cite{BuDoMoRoSt}. 

Instead of reproducing here 
the preceding statements about the asymptotic quantum corrections
of higher moments of certain specific observables
by explicit computations, we sketch  
a quite general related result. Namely, 
given any subalgebra ${\cal C}$ of local observables, stable 
under translations and irreducible in the vacuum sector,
we shall show that the 
quantum corrections cannot vanish for all elements of ${\cal C}$.
In particular, the algebra 
generated by the charge density and the magnetic 
field can be shown to satisfy these conditions. So there must be 
polynomials in these observables giving rise 
to non--trivial quantum corrections, in accordance with the 
computational results.

For the proof of the above statement, we consider the  
maps $\delta_j$, $j=1,2,3$, from 
${\cal C}$ into the algebra of all
local observables, given by 
\be
\delta_j (C) \equiv i \! \int \! dz \, h(z_0) \, [ E_j (z), C], \quad 
C \in {\cal C}.
\ee
Because of locality, these maps are well defined. Assuming that 
the quantum corrections of all elements of ${\cal C}$ vanish yields
(since ${\cal C}$ is invariant under translations) 
\begin{eqnarray}
& 0 = \int \! dz \, h(z_0) \, (\Omega, [ A_j (z), \partial_0 \, C] 
\, \Omega) =  - \int \! dz \, h(z_0) \, (\Omega, [ \partial_0 A_j (z), 
C] \, \Omega) & \nonumber \\
& = - \int \! dz \, h(z_0) \, ( \Omega, [E_j (z), C] 
\, \Omega), &
\end{eqnarray} 
where the third equality is a consequence of the fact that 
$\partial_j A_0 (z)$ does not contribute to the integral because
of locality. Thus 
\be
(\Omega, \delta_j (C) \, \Omega) = 0, \quad  C \in {\cal C},
\ee
so one can consistently define Hermitian operators $Q_j$ in the
vacuum sector, setting
\be
Q_j \, C \Omega \equiv \delta_j (C) \, \Omega, \quad  C \in {\cal C}.
\ee
Moreover, it follows from the generalization of a famous result 
of Coleman, cf.\ \cite{BrBuGl}, that these operators are 
(combinations of) constants of motion. Thus we conclude
that the components $E_j$ of the electric field are subject to a 
conservation law. This is indeed so for the free electromagnetic field,
where $\partial^\nu F_{\nu \mu} = 0$, but clearly not so in quantum
electrodynamics. So non--trivial asymptotic quantum corrections inevitably 
appear in this case for some elements of ${\cal C}$. 

We conclude this section by noting that 
the existence of some subalgebra ${\cal C}$ 
of observables where all quantum corrections vanish 
would be a prerequisite for applying
the methods outlined in \cite{BuDoMoRoSt}.
So an analysis of the superselection structure and statistics of 
the electrically charged states considered here  
cannot be carried out along these lines.
%%%%%%%%%%%%%%%%%%%%%%%%%%%%%%%%%%%%%%%%%%%%%%%%%%%%%%%%%%%%%%%%%%%%%%%%%%%%
%%%%%%%%%%%%%%%%%%%%%%%%%%%%%%%%%%%%%%%%%%%%%%%%%%%%%%%%%%%%%%%%%%%%%%%%%%%%
%%%%%%%%%%%%%%%%%%%%%%%%%%%%%%%%% SECTION 4 %%%%%%%%%%%%%%%%%%%%%%%%%%%%%%%%
%%%%%%%%%%%%%%%%%%%%%%%%%%%%%%%%%%%%%%%%%%%%%%%%%%%%%%%%%%%%%%%%%%%%%%%%%%%%
%%%%%%444444444444444444444444444444444444444444444444444%%%%%%%%
%%%%%%%%%%%%%%%%%%%%%%%%%%%%%%%%%%%%%%%%%%%%%%%%%%%%%%%%%%%%%%%%%
\section{Conclusions}
The quantum delocalization of the electric charge, established in the 
present investigation, is due to a combination of quantum effects and
the influences of interaction. The subtle interplay between these ingredients
causes a Coulomb like spreading of (higher moments of) the charge density,
present neither in the interacting classical theory nor in 
quantum field theory if the interaction is turned off. In fact, the 
amplitudes of these long range contributions are suppressed by powers
of the fine structure constant and consequently are extremely small. 
It may therefore be difficult, if not impossible to establish their existence 
experimentally.

On the theoretical side, however, this delocalization
of the charge means that the 
notion of charge support is fraught with conceptual difficulties. 
For the qualitative picture of a well--localized
charge distribution does not have a clearcut mathematical 
counterpart. This fact gives rise to complications in the 
discussion of the statistics and fusion structure of electrically 
charged states, where localization properties matter. In particular,
the general methods outlined in \cite{BuDoMoRoSt} cannot be applied
in this case.

Although it seems impossible to discriminate the electric charge and 
field support, the coarser notion of causal support is
still meaningful. 
We recall that a state is causally supported in a region if 
it can be generated from the vacuum by some physical isometric 
operation localized there. 
The causal support of states carrying electric charge
is clearly non--compact, but it can be confined to an 
arbitrary spatial cone.
For there are charged physical fields with gauge fixing functions
having support in such cones \cite{St,Bu} and one can proceed from them
to corresponding charged isometries by a process of polar decomposition,
described in Section 3.

As pointed out in \cite{Bu}, such cone--like localized operators could be
the starting point for a systematic analysis of the statistics and fusion
structure of the superselection sectors in theories with electromagnetic
forces. This would require, however, a better understanding of the relation
between operators localized in different cones. In particular, it 
would be necessary to show that these operators are related  
by suitable limits of local observables which merely describe different  
configurations of low energy photons, in analogy to the situation 
discussed in \cite{Ku}. It should be possible to provide
evidence to this effect by perturbative methods similar to those 
used in the present investigation.
 
%%%%%%%%%%%%%%%%%%%%%%%%%%%%%%%%%%%%%%%%%%%%%%%%%%%%%%%%%%%%%%%%%%%%%%%%%%%
%%%%%%%%%%%%%%%%%%%%%%%%%%%%%%%%%%%%%%%%%%%%%%%%%%%%%%%%%%%%%%%%%%%%%%%%%%%
%%%%%%%%%%%%%%%%%%%%%%%%%%%%%%%%%%% APPENDIX
%%%%%%%%%%%%%%%%%%%%%%%%%%%%%%%%%%% %%%%%%%%%%%%%%%%%%%%%%%%%%%%%%
%%%%%%%%%%%%%%%%%%%%%%%%%%%%%%%%%%%%%%%%%%%%%%%%%%%%%%%%%%%%%%%%%%%%%%%%%%
%%%%%%%%%%%%%%%%%%%%%%%%%%%%%%%%%%%%%%%%%%%%%%%%%%%%%%%%%%%%%%%%%%%%%%%%%%
\vspace*{5mm}
{\noindent \Large \bf Appendix} \\[3mm]
\setcounter{section}{1}
\setcounter{equation}{0}
\renewcommand{\theequation}{\Alph{section}.\arabic{equation}}
\ni In this appendix we establish the existence of certain specific 
(generalized) solutions $y \rightarrow \gmy$ of the equation
\be \label{A.1}
\partial^\mu\gmy = - \delta(y-x),
\ee
where $x = (\x,x_0)$ lies in the bounded region $|\x| < R$,
$|x_0|<T$. To avoid 
overburdening the notation, we will not indicate 
the dependence of these functions on $x$ in the following. 

The desired solutions can be decomposed into 
$g_\mu = g_\mu^I + g_\mu^{\roii}$, where $g_\mu^I$, $g_\mu^{\roii}$
have the following specific properties. The function $g_\mu^I$ has
compact support in the region
$|\y | < 4 R, \ |y_0| < T $, for $x$ varying within the above limitations, 
and is given by
\be \label{A.2}
g_0^I(y) = 0, \quad g_i^I (y) = - \ovp 
(y_i - x_i) \, |\y - \x|^{-3} \, \delta(y_0 - x_0), \ i=1,2,3 
\ee
for $y \in {\cal R}_I  \equiv \{ z : |\z | < R \}$.
The function
$g_\mu^{\roii}$ does not depend on the choice of $x$,
vanishes for $|\y | < 3 R$, and is, for  
$y \in {\cal R}_\roii  \equiv \{ z : |\z | > 4R \}$, given by
\be \label{A.3}
g_0^\roii(y) = 0, \quad g_i^\roii (y) = - \ovp y_i \,  |\y|^{-3} \, h(y_0), 
\ i=1,2,3,
\ee
where $h$ is a test function with support in $I_T \equiv (-T,T)$ and 
$\int \! dy_0 \, h(y_0) =1$. 
Note that $g_\mu^I$, $g_\mu^{\roii}$ are
local solutions of equation (\ref{A.1}) in the regions ${\cal R}_I$
and ${\cal R}_\roii$, respectively. 

For the proof that a corresponding global interpolating solution exists, 
we make an ansatz of the form
\be \label{A.4}
g_i (\y, y_0) = - f_i(\y) \, k(\y, y_0), \ \ i=1,2,3.  
\ee
With this ansatz, Eq.\ (\ref{A.1}) is 
clearly satisfied if
\begin{eqnarray}
& \partial^i f_i(\y) = \delta (\y-\x), & \label{A.5} \\ 
& \delta(\y-\x) \, k(y) = \delta(y-x) & \label{A.6} 
\end{eqnarray}
and $g_0$ is a solution of 
\be 
\partial^0 g_0 (y) = f_i (\y) \; \partial^i k(\y, y_0).
\label{A.7} 
\ee
If $k$ is chosen to have compact support in 
$I_T$ with respect to $y_0$ and
\be
\int dy_0 \,  k(\y,y_0)  = 1  \ \ \mbox{for all} \ \ \y,
  \label{A.8} 
\ee
the solutions $g_\mu$ have compact support
as well. In fact, integrating the
right hand side of Eq.\ (\ref{A.7}) 
with respect to $y_0$ yields $0$ for all $\y$, 
hence Eq.\ (\ref{A.7})
has a (unique) solution $g_0$ which has 
compact support in $I_T$ with respect to $y_0$. 

The given form of the functions $g_\mu^I$, $g_\mu^{\roii}$   
in the regions ${\cal R}_I$ and 
${\cal R}_{\roii}$, respectively, is consistent with the above ansatz.
More concretely, $g_\mu^I$ corresponds to the choice 
\be \label{A.9}
f_i^I (\y) = \ovp (y_i - x_i) \, |\y - \x|^{-3}, \quad 
k^I (y) = \delta (y_0-x_0) 
\ee
in Eqs.\ (\ref{A.4}), (\ref{A.7}) 
and similarly $g_\mu^{\roii}$ is fixed by   
\be \label{A.10}
f_i^{\roii} (\y) = \ovp  y_i \, |\y |^{-3},   \quad
k^{\roii} (y) = h (y_0). 
\ee
To interpolate these local solutions 
and solve Eq.\ (\ref{A.1}) on all of $\RR^4$,
we first construct functions $f_i$ which 
satisfy Eq.\ (\ref{A.5}) and 
coincide with $f_i^I$ for $|\y| < R$ as well as   
with $f_i^{\roii}$ for $|\y| > 3R$, $i=1,2,3$.  
These functions are obtained as components of the  
electric field corresponding to the retarded solution
of Maxwell's equations for a suitable current. The 
corresponding charge density is given by  
\be \label{A.11}
j_0 (\z,z_0) \equiv \delta (\z - \x(z_0)),
\ee
where $z_0 \rightarrow \x(z_0)$ is assumed to be a smooth
function which is   
equal to $0$ for $z_0 < - 3R $, 
equal to $\x$ for $ z_0 > - 2R$, and which satisfies  
$| d \x(z_0)/dz_0| < 1$. As $|\x| < R$, 
such a function exists.
The spatial components of the current are given by
\be \label{A.12}
j_i (\z,z_0) \equiv \delta (\z - \x(z_0)) \, dx_i (z_0)/dz_0,
\ i=1,2,3,
\ee
so the resulting current 
$j_\mu$ is conserved. The corresponding retarded potential is 
\be \label{A.13}
a_\mu (y) = \int d\z \; |\z-\y|^{-1} \, 
   j_\mu (\z, y_0 - |\z-\y|). 
\ee
Taking into account the spacetime features of the 
current, a moment's reflection shows that the 
associated electric field 
\be \label{A.14}
f_i (\y) \equiv \partial_0  a_i(\y,0) - \partial_i a_0(\y,0), \ \
i=1,2,3,
\ee
satisfies Eq.\ (\ref{A.5}) and coincides with
$f_i^I$ and $f_i^{\roii}$ in the regions 
$|\y| < R$ and  $|\y| > 3R$, respectively. 

It remains to show that there is a function $y
\rightarrow k (y)$ with properties specified above  
interpolating between $k^I$ and $k^{\roii}$.
To this end we pick a smooth function $\y \rightarrow l (\y)$
which is equal to $0$ for $|\y| < 3 R$ and  equal to 
$1$ for $|\y| > 4 R$ and put
\be \label{A.15}
k(y) \equiv (1 - l(\y))  \, \delta (y_0 -x_0) 
+ l (\y) \, h (y_0)
\ee 
with $h$ as above. Taking into account 
the restrictions on $x$, it follows that 
$k$ has
support in $I_T$  with respect to $y_0$ 
and satisfies Eqs.\ (\ref{A.6}) and (\ref{A.8}). 

By making use of Eqs.\ (\ref{A.4}) and (\ref{A.7}) with 
$f_i$ and $k$ defined as above, 
we arrive at solutions $g_\mu$ of (\ref{A.1}).
The corresponding functions $g_\mu^I$, $g_\mu^{\roii}$ are
given by 
\begin{eqnarray}
& g_0^I(y) \equiv - 
f_i(\y) \, \partial^i l(\y) \, (H(y_0) - \theta(y_0 - x_0)),
& \nonumber \\ 
& g_i^I(y) \equiv - f_i(\y) \, (1 - l(\y))  \, \delta (y_0 -x_0), \ 
i=1,2,3, & \label{A.16}  
\end{eqnarray}
where $H$ is the primitive of $h$ vanishing for $y_0 < - T$, and 
\be \label{A.17}
 g_0^\roii(y) \equiv 0, \quad
g_i^\roii(y) \equiv  - \ovp  y_i \, |\y |^{-3} \, l(\y) \, h(y_0), \   
i=1,2,3.
\ee
Bearing in mind the properties of the $f_i$ and $l$, it follows that 
these functions have all
the desired features stated above.

\vspace*{5mm}
\ni {\Large \bf Acknowledgements}\\[2mm]
Our collaboration was made possible by financial support from
the CNR, INFN, the Scuola Normale Superiore  
and the Universtit\`a di Roma ``La Sapienza'',
which we gratefully acknowledge. 
One of us (DB) is also grateful   for
hospitality shown to him at the 
Scuola Normale Superiore and the  
Universit\`a di Roma ``La Sapienza'' and ``Tor Vergata''. 

%%%%%%%%%%%%%%%%%%%%%%%%


\begin{thebibliography}{99}
\bibitem{Fr} J.\ Fr\"{o}hlich, Commun.\ Math.\ Phys.\ {\bf 66} (1979) 223
%%CITATION = CMPHA,66,223;%% 
\bibitem{BuDoMoRoSt} D.\ Buchholz, S.\ Doplicher, G.\ Morchio, J.E.\
Roberts and F.\ Strocchi, pp.\ 647--660 in: {\it Operator Algebras and 
Quantum Field Theory: Rome 1996}, International Press 1997
%%CITATION =  hep-th/9705089;%%
\bibitem{DoHaRo} S.\ Doplicher, R.\ Haag and J.E.\ Roberts, Commun.\ Math.\
Phys.\ {\bf 23} (1971) 199 
%%CITATION = CMPHA,23,199;%%  
\ and \ Commun.\ Math.\ Phys.\ {\bf 35} (1974)  49 
%%CITATION = CMPHA,35,49;%% 
\bibitem{DoRo} S.\ Doplicher and J.E.\ Roberts, Commun.\ Math.\ Phys.\ 
{\bf 131} (1990) 51 
%%CITATION = CMPHA,131,51;%% 
\bibitem{Sw} J.\ Swieca, Nuovo Cimento {\bf 52A} (1967) 242, 
%%CITATION = NUCIA,52A,242;%% 
\bibitem{St} O.\ Steinmann, Ann.\ Phys.\  {\bf 157} (1984) 232
%%CITATION = APNYA,157,232;%%
 \ and \ {\it Perturbative Quantum Electrodynamics and Axiomatic
Field Theory}, Springer 2000
%%CITATION = NONE;%%
\bibitem{KlMa} S.\ Klainermann and M.\ Machedon, Duke Math.\ Jour.\ {\bf
74} (1994) 19 
%%CITATION = DUMJA,74,19;%%
\bibitem{PaStVe} C.\ Parenti, F.\ Strocchi and G.\ Velo, Commun.\ Math.\
Phys.\ {\bf 53} (1997) 65 
%%CITATION = CMPHA,53,65;%% 
\bibitem{FlSiTa1} M.\ Flato, J.C.H.\ Simon and E.\ Taffin, Rev.\ Math.\
Phys.\ {\bf 6} (1994) 1071
%%CITATION = RMPHE,6,1071;%% 
\bibitem{FlSiTa2} M.\ Flato, J.C.H.\ Simon and E.\ Taffin, 
Memoirs of Am.\ Math.\ Soc. Vol.\ 606, 1997
%%CITATION = MAMCA,606;%%
\bibitem{BlSe} P.\ Blanchard and R.\ Seneor, Ann.\ Inst.\ H.\ Poincar\'e
A{\bf 23} (1975) 147
%%CITATION = AHPAA,23,147;%%
\bibitem{FeHuRoWr} J.S.\ Feldman, T.R.\ Hurd, L.\ Rosen and J.D.\ Wright,
{\it QED: A proof of renormalizability}, Springer 1988
%%CITATION = NONE;%%
\bibitem{KeKo} G.\ Keller and C.\ Kopper, Commun.\ Math.\ Phys.
{\bf 176} (1996) 193
%%CITATION = CMPHA,176,193;%%
\bibitem{Sy} K. Symanzik, Lectures on Lagrangean Quantum Field
Theory, DESY report T-71/1
%%CITATION = NONE;%%
\bibitem{Sch}  G.\ Scharf, {\it Finite quantum electrodynamics}, 2nd
edition, Springer 1995
%%CITATION = NONE;%%
\bibitem{StWi} F. Strocchi and A.S. Wightman, J.\ Math.\ Phys.\ {\bf
15} 2198 (1974)
%%CITATION =  JMAPA,15,2198;%%
\bibitem{DuFr} M.\ D\"utsch and K.\ Fredenhagen, Commun.\ Math.\ Phys.\
{\bf 203} (1999) 71 
%%CITATION = CMPHA,203,71;%%
\bibitem{Bo} H.--J.\ Borchers, Commun.\ Math.\ Phys.\ {\bf 4} (1967) 315 
%%CITATION = CMPHA,4,315;%%
\bibitem{Bo2} H.--J.\ Borchers, Nuovo Cimento {\bf 33} (1964) 1600
%%CITATION = NUCIA,33,1600;%%
\bibitem{ArHeRu} H.\ Araki, K.\ Hepp and D.\ Ruelle, Helv.\ Phys.\ Acta
{\bf 35} (1962) 164 
%%CITATION =  HPACA,35,164;%%
\bibitem{Re} M.\ Requardt, Commun.\ Math.\ Phys.\ {\bf 50} (1976) 259 
%%CITATION = CMPHA,50,259;%%
\bibitem{MoSt} G.\ Morchio and F.\ Strocchi, Ann.\ Phys.\ {\bf 168} (1986) 27
%%CITATION = APNYA,168,27;%%
\bibitem{JaRo} J.M.\ Jauch and F.\ Rohrlich, {\it The theory of 
photons and electrons}, 2nd edition, Springer 1976
%%CITATION = NONE;%% 
\bibitem{BrBuGl} J.\ Bros, D.\ Buchholz and V.\ Glaser,
Commun.\ Math.\ Phys.\ {\bf 50} (1976) 11
%%CITATION = CMPHA,50,11;%%
\bibitem{Bu} D.\ Buchholz, Commun.\ Math.\ Phys.\ {\bf 85} (1982) 49
%%CITATION = CMPHA,85,49;%%
\bibitem{Ku} W.\ Kunhardt, J.\ Math.\ Phys.\ {\bf 39} (1998) 6353
%%CITATION = JMAPA,39,6353;%%
\end{thebibliography}
\end{document}